\documentclass[aps,prb,twocolumn,floatfix,superscriptaddress]{revtex4}
\usepackage{bm}
\usepackage{epsf}
\usepackage{amssymb}
\usepackage{amsmath}
\usepackage{graphicx}
\usepackage{rotating}
\usepackage{epsfig}
\usepackage{psfrag}
\usepackage{amsmath}
\usepackage{hyperref}
\usepackage{subfigure}
\usepackage{bm}
\usepackage{color}

\renewcommand{\dag}{\dagger}

\renewcommand{\dag}{\dagger}

\def\mb{\mathbf}
\def\mr{\mathrm}
\def\mc{\mathcal}

\newcommand{\be}{\begin{equation}}
\newcommand{\ee}{\end{equation}}
\newcommand{\ba}{\begin{eqnarray}}
\newcommand{\ea}{\end{eqnarray}}

\newcommand {\apgt} {\ {\raise-.5ex\hbox{$\buildrel>\over\sim$}}\ }
\newcommand {\aplt} {\ {\raise-.5ex\hbox{$\buildrel<\over\sim$}}\ }

\begin{document}

\title{Enhanced stability of skyrmions in two-dimensional chiral magnets with Rashba spin-orbit coupling}

\author{Sumilan Banerjee}
\affiliation{Department of Physics, The Ohio State University, Columbus, Ohio, 43210}
\author{James Rowland}
\affiliation{Department of Physics, The Ohio State University, Columbus, Ohio, 43210}
\author{Onur Erten}
\affiliation{Department of Physics, The Ohio State University, Columbus, Ohio, 43210}
\affiliation{Department of Physics and Astronomy, Rutgers University,   Piscataway, New Jersey, 08854}
\author{Mohit Randeria}
\affiliation{Department of Physics, The Ohio State University, Columbus, Ohio, 43210}

\date{\today}

\begin{abstract}
Recent developments have led to an explosion of activity on 
skyrmions in three-dimensional (3D) chiral magnets. 
Experiments have directly probed these topological spin textures, revealed their non-trivial properties,
and led to suggestions for novel applications. 
However, in 3D the skyrmion crystal phase is observed only in a narrow region of the temperature-field phase diagram.
We show here, using a general analysis based on symmetry, that skyrmions are much more readily stabilized in two-dimensional (2D)
systems with Rashba spin-orbit coupling.
This enhanced stability arises from the competition between field and easy-plane magnetic anisotropy,
and results in a nontrivial structure in the topological charge density in the core of the skyrmions.
We further show that, in a variety of microscopic models
for magnetic exchange, the required easy-plane anisotropy naturally 
arises from the same spin-orbit coupling that is responsible for the chiral 
Dzyaloshinskii-Moriya interactions.
Our results are of particular interest for 2D materials like thin films, surfaces and oxide interfaces, 
where broken surface inversion symmetry and Rashba spin-orbit coupling 
leads naturally to chiral exchange and easy-plane compass anisotropy.
Our theory gives a clear direction for experimental studies of 2D magnetic materials
to stabilize skyrmions over a large range of magnetic fields down to $T\!=\!0$.
\end{abstract}

\maketitle

\section{Introduction}
Skyrmions first arose in the study of hadrons in high energy physics~\cite{NuclPhysSkyrme1962}, but
these topological objects have proved to be central in the study of chiral magnets~\cite{NatureNanoNagaosa2013,JMMMBogdanov1994,NatureRossler2006}, in addition 
to a variety of other condensed matter systems, including quantum Hall effect~\cite{PRBSondhi1993,PRLBarrett1995,PRLSchmeller1995}
and ultra cold atoms~\cite{PRLHo1998,PRLCole2012,PRLRadi2012}.
There has been tremendous progress in establishing exotic skyrmion crystal (SkX) phases,
using neutrons~\cite{ScienceMuhlbauer2009} and Lorentz transmission electron 
microscopy~\cite{NatureYu2010}, in a variety of magnetic materials 
that lack bulk inversion symmetry, ranging from metallic helimagnets like MnSi~\cite{NatureNanoNagaosa2013,ScienceMuhlbauer2009}  
to insulating multiferroics~\cite{ScienceSeki2012}. Skyrmions lead to unusual transport properties in metals like the topological Hall 
effect~\cite{PRLNeubauer2009,PRLLee2009,NatureRitz2013,PRLFranz2014}, they may be related to 
non-Fermi liquid behavior~\cite{NaturePfleiderer2001,NatureLeyraud2003,ArXivVishwanath2013},
and they could have potential applications in spintronics~\cite{NatureNanoNagaosa2013,ScienceJonietz2010,PRLLin2013,PRBKnoester2014}.

\begin{figure}
\begin{center}
\includegraphics[width=7.5cm]{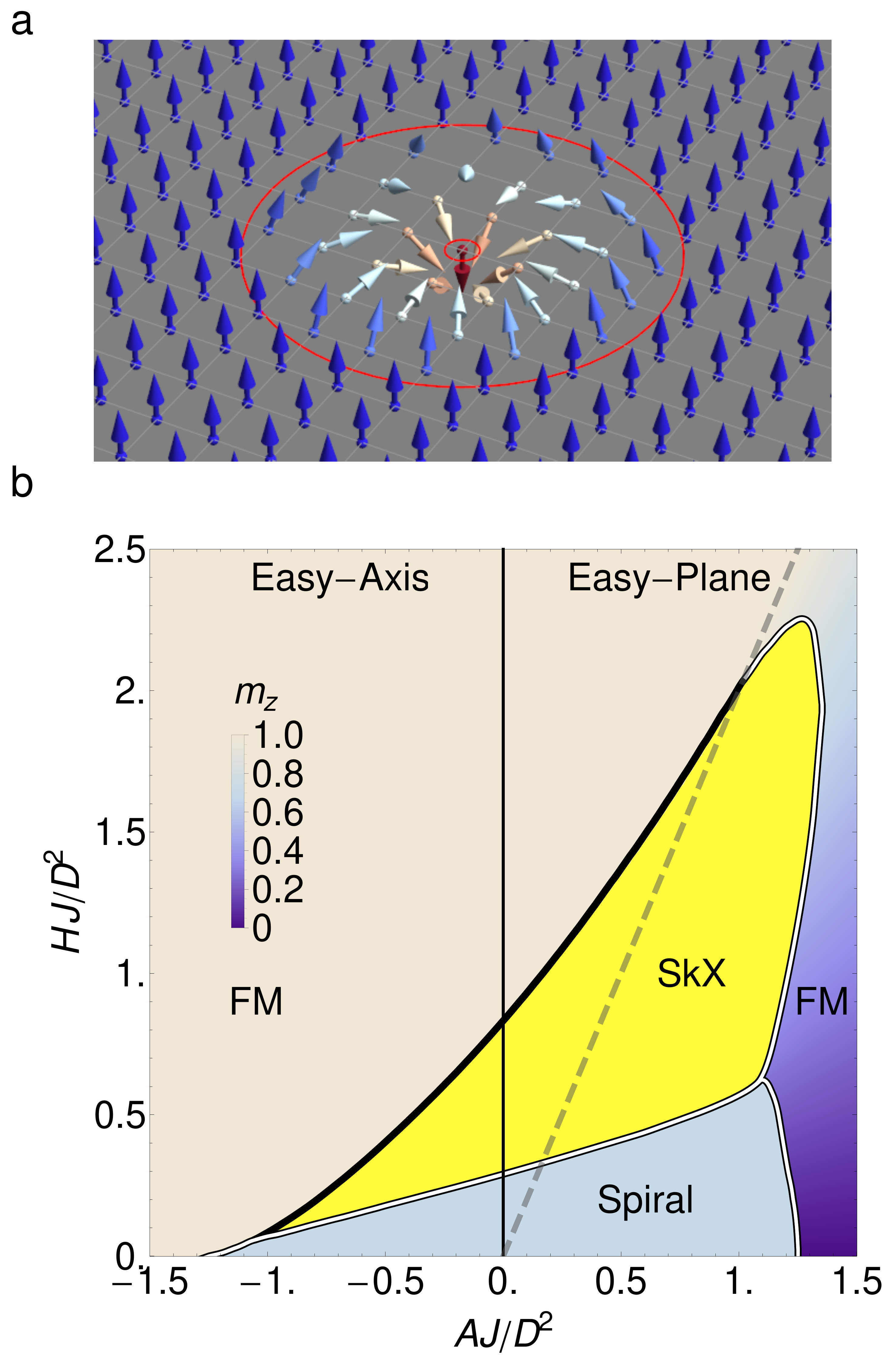}
\end{center}
\caption{{\bf Skyrmion and $T\!=\!0$ phase diagram} (a) A skyrmion configuration. 
(b) The anisotropy-field phase diagram with ferromagnetic (FM), spiral and skyrmion crystal (SkX) phases 
for $D/J=0.01$ and $A_cJ/D^2=1/2$ with $A = A_c + A_s$. 
Double lines denote first order transitions, while the single line is an unusual first order transition with a divergent length scale; see text.
The dashed line $H=2A$ separates the out-of-plane FM from the tilted FM. $m^z$ for the FM is shown in the color bar.
Results are obtained from a circular-cell variational calculation.
}
\label{fig.PhaseDiagram}
\end{figure}

Spin-orbit coupling (SOC) in magnetic systems without inversion gives rise to the 
chiral Dzyaloshinskii-Moriya (DM) \cite{Dzyaloshinskii1958,PRMoriya1960} interaction
$\mb{D}_{ij}\!\cdot\!(\mb{S}_i\times \mb{S}_{j})$. This competes with the usual $\mb{S}_i\!\cdot\!\mb{S}_{j}$ exchange
to produce spatially modulated states like spirals and SkX.

The 2D case is particularly interesting. Even in materials that break bulk inversion, thin films show enhanced 
stability~\cite{NatureMatYu2011,PRBButenko2010} of skyrmion phases, persisting down to lower temperatures.
Inversion is necessarily broken in 2D systems on a substrate or at an interface, and this too may lead to textures 
arising from DM interactions. Spin-polarized STM~\cite{NatureBode2007,ScienceRomming2013} has observed such textures on magnetic monolayers 
deposited on non-magnetic metals with large SOC. 

Recently, there have been tantalizing hints of magnetism at oxide interfaces like 
LaAlO$_3$/SrTiO$_3$~\cite{NatMatBrinkman2007,NatPhysLi2011,NatPhysBert2011,NatMatLee2013} and 
GdTiO$_3$/SrTiO$_3$\cite{PRXMoetakef2012}. The 2D electron gas at the interface between 
two insulating oxides has a large and gate-tunable Rashba SOC~\cite{PRLCaviglia2010}.
We have proposed~\cite{NaturePhysBanerjee2013} that broken surface inversion and Rashba SOC at oxide interfaces
necessarily leads to chiral magnetic interactions, thus leading to phases with 
spin textures~\cite{NaturePhysBanerjee2013,PRLBalents2014}.

%
%
%
With this motivation we investigate 2D chiral magnets 
with broken inversion in the $z$-direction. Microscopically, this leads to Rashba SOC.
General symmetry considerations imply that the form of the free energy
for broken surface inversion (see eq.~(\ref{Rashba-DM})) is quite different from that in the
usually studied case of non-centrosymmetric materials
with broken bulk inversion.

Our results are summarized in the $T = 0$ phase diagram in Fig.~\ref{fig.PhaseDiagram}
as a function of perpendicular magnetic field $H$ and anisotropy $A$.
For easy-axis anisotropy ($A < 0$), our 2D results with broken $z$-inversion turn out to be essentially the same
as those for the 3D problem with broken bulk inversion~\cite{JMMMBogdanov1994,PRBButenko2010}.
The easy-plane regime ($A > 0$) in the 2D Rashba case leads to a surprise: we find an unexpectedly large stable SkX phase.
Skyrmions not only gain DM energy, but are also an excellent compromise between the field and easy-plane anisotropy. 
Moreover, we show that the skyrmions have a nontrivial spatial variation of their topological charge density
(see Fig.~\ref{fig.skyrmion-structure}) for $A>0$. 
 
Can such easy-plane anisotropy of the required strength arise naturally in real materials?
We present a microscopic analysis of three exchange mechanisms 
-- superexchange in Mott insulators, and double exchange and Ruderman-Kittel-Kasuya-Yosida (RKKY) interaction in metals --
and show that the same SOC that gives rise to the DM interaction $D$ also 
leads to an easy-plane compass anisotropy $A_c$. The compass term is usually ignored 
since it is higher order in SOC than DM. We show, however, that its contribution to the energy 
is comparable to that of DM, with $A_c|J|/D^2 \simeq1/2$ for all three mechanisms, where $J$ is
the exchange coupling.
This striking fact seems not to have been clearly recognized earlier, possibly because these 
microscopic mechanisms have been discussed in widely different contexts 
using different notation and normalizations.
We also discuss how additional single-ion anisotropies enter the analysis.

Our results should serve as a guide for material parameters
of 2D chiral magnets such that a large SkX region can be probed experimentally.
These results are of particular relevance to magnetism at oxide interfaces
as discussed above. 
We should also emphasize that our 2D results are not
necessarily restricted to monolayers. We discuss the case of quasi-2D materials
in Section V.

\section{Ginzburg-Landau Theory} 
The continuum free-energy functional ${F}[\mb{m}]=\int d^2r \mc{F}(\mb{m})$
for the local magnetization $\mb{m}(\mb{r})$ of a 2D chiral magnet in an applied field $\mb{H}$ is given by
\begin{eqnarray}
 \mc{F}=\mc{F}_\mr{iso}(\mb{m})+\mc{F}_\mr{DM}(\mb{m})+\mc{F}_\mr{aniso}(\mb{m})-\mb{H}.\mb{m}. 
 \label{eq.GL}
 \end{eqnarray}  
The isotropic term ($\alpha=x,y,z$)
\begin{eqnarray}
\mc{F}_\mr{iso}=\mc{F}_0(\mb{m})+(J/2)\sum_\alpha (\nabla m^\alpha)^2
\end{eqnarray}
consists of $\mc{F}_0$ that determines the magnitude of $\mb{m}$
and a stiffness $J$ that controls the gradient energy. 
Microscopically the stiffness is determined by the ferromagnetic exchange coupling.
At $T\!=\!0$ we replace
$\mc{F}_0$ with the constraint $\mb{m}^2(\mb{r})\!=\!1$.  
Rashba SOC, arising from broken $z$-inversion, leads to the DM term
\begin{eqnarray}
\mc{F}_\mr{DM}&=&-D[(m^z\partial_xm^x-m^x\partial_xm^z)\nonumber
\\
&&-(m^y\partial_ym^z-m^z\partial_ym^y)].
\label{Rashba-DM}
\end{eqnarray} 
We will see below that this leads to ``hedgehog''-like skyrmions (Fig.~\ref{fig.PhaseDiagram}(a)).
This form of $\mc{F}_\mr{DM}$ is dictated by the 
DM vector $\mb{D}_{ij} \sim \widehat{\mb{z}}\times\widehat{\mb{r}}_{ij}$
for Rashba SOC. 
In contrast, broken bulk inversion with $\mb{D}_{ij} \sim \widehat{\mb{r}}_{ij}$ gives rise to the 
more familiar DM term $\mb{m}\cdot(\nabla\!\times\!\mb{m})$ that leads to ``vortex''-like skyrmions.
\footnote{
We note that the $\mb{m}\cdot(\nabla\!\times\!\mb{m})$ DM interaction can be transformed to $\mc{F}_\mr{DM}$ of eq.~(\ref{Rashba-DM})
by a global $\pi/2$-rotation of $\mb{m}$ about the $z$ axis. We will not exploit this transformation here 
since we focus only on the Rashba SOC in this paper. See, however, Section V, where we comment on
the case where both surface and bulk inversion is broken.}
%
%

Rashba SOC also leads to the anisotropy term  
\begin{eqnarray}
\mc{F}_\mr{aniso}&=&(A_c/2)[(\partial_xm^y)^2+(\partial_y m^x)^2]\nonumber
\\
&-& A_c[(m^x)^2+(m^y)^2]+A_s(m^z)^2.
\end{eqnarray}
The $A_c\!>\!0$ ``compass" terms give rise to  easy-plane anisotropy, while the single-ion
$A_s$ term can be either easy-axis ($A_s\!<\!0$) or easy-plane ($A_s\!>\!0$).
We define length in units of lattice spacing $a$ so that $J$, $D$, $A_c$ and $A_s$ all have dimensions of energy.   

While the form of the free energy \eqref{eq.GL} follows from symmetry, the microscopic analysis 
(described in Section IV) gives insight into the relative strengths of various terms.
The origin of the DM and compass terms lies in Rashba SOC, whose strength $\lambda\ll t$, the hopping, in materials 
of interest. Thus we obtain a hierarchy of scales with the exchange $J\gg D\!\sim\!J(\lambda/t)\gg A_c\!\sim\!J(\lambda/t)^2$. 
Naively one might expect the compass term to be unimportant, however its contribution to the energy
${\cal O}(A_c)$ is comparable to that of the DM term ${\cal O}(D^2/J)$. While the DM term
is linear in the wave-vector $\mb{q}$ of a spin configuration, its energy must be ${\cal O}(q^2)$.
Thus compass anisotropy, usually ignored in the literature, must be taken into account whenever the DM term
is important.

We show below that $A_c J/D^2\simeq 1/2$ for a wide variety of exchange mechanisms
independent of whether the system is a metal or an insulator.  
We also discuss the origin and strength of the single-ion $A_s$ term. Note
that the {\it effective anisotropy} in model \eqref{eq.GL} is governed by 
$A = A_c + A_s$, which is easy-axis  for $A\!<\!0$ and easy-plane for $A\!>\!0$.

\section{Phase Diagram} 
We begin by examining the $T\!=\!0$ phase diagram
for fixed $D\ll J$ as function of magnetic field $\mb{H}=H\hat{z}$
and the dimensionless anisotropy $AJ/D^2$, which we explore
by varying $A_s$ with $A_cJ/D^2=1/2$. 
We look for variational solutions using 
analytical and numerical approaches.  Here we focus on the SkX phase;
the ferromagnetic (FM) and spiral phases
are discussed in Appendix \ref{App:VarCaln}. 

A skyrmion\cite{NatureNanoNagaosa2013} is a spin-texture with 
a quantized topological charge 
$q=(4\pi)^{-1}\int d^2{\bf r} \ \hat{\mb{m}}\cdot(\partial_x\hat{\mb{m}}\times\partial_y\hat{\mb{m}})$,
which is restricted to be an integer. For example, the $q=-1$ skyrmion in Fig.~\ref{fig.PhaseDiagram}(a) is a 
smooth spin configuration with the topological constraint
that the central spin points down while all the spins at the boundary point up.

The SkX state is a periodic array of skyrmions,
often described by multiple-$\mb{Q}$ spiral condensation \cite{PRLBinz2006,PRBYi2009}. 
We use an `optimal unit-cell' approach, similar to ref.~\onlinecite{JMMMBogdanov1994}, 
where we impose the topological constraint for the center and boundary spins
within a unit cell.
We then find the optimal configuration within a single cell, whose size $R$ is also determined variationally.

We describe the results from a `circular-cell' ansatz, which leads to an effectively 1D (radial)
problem. This is computationally much simpler than the full 2D conjugate-gradient minimization of \eqref{eq.GL}.
The 2D and 1D methods lead to essentially identical phase diagrams; 
see Appendix ~\ref{App:2DMin}. Here,
we take a skyrmion configuration 
\begin{eqnarray}
\mb{m}_{\rm skyrmion}(\mb{r})&=&\sin\theta(r)\hat{\mb{r}}+\cos\theta(r)\hat{\mb{z}} \label{eq.Skyrmion}
\end{eqnarray}
in a circular cell of radius $R$, with the topological constraint $\theta(0)=\pi$ and $\theta(R)=0$. 
We minimize the energy \eqref{eq.GL} with $\theta(r)$ and the cell radius $R$ as variational parameters.
We construct the SkX by an hexagonal packing of the optimal circular cells 
and recalculate the energy with up spins filling the space between the circles.

%
As a first step, we use the linear ansatz~\cite{JMMMBogdanov1994} $\theta(r)=\pi(1-r/R)$
with skyrmion size $R$, a simple approximation that has the great virtue of being analytically tractable.
The resulting phase diagram is shown in the Appendix (see dotted lines in Fig.~\ref{fig.PhaseDiagram_SquareCell})
rather than in the main text, so as not to clutter up Fig.~\ref{fig.PhaseDiagram}(b).
We note here that this very simple approximation  already 
gives us our first glimpse of the large SkX phase for easy-plane anisotropy,
despite the fact that it greatly underestimates the stability of the SkX phase.

Next we obtain the phase diagram in Fig.~\ref{fig.PhaseDiagram}(b)
by numerical minimization using the more general form of eq.~\eqref{eq.Skyrmion} and discretizing $\theta(r)$ on a 1D grid. 
This confirms the qualitative observations from the linear approximation and 
yields an even larger SkX phase on the easy-plane side.
Our 2D square cell calculations essentially reproduce the same phase diagram (see Fig.~\ref{fig.PhaseDiagram_SquareCell}).

\subsection{Easy-plane vs. easy-axis anisotropy} 
Our results for the 2D phase diagram in the easy-axis region ($A < 0$)
is much the same as previous 3D studies~\cite{JMMMBogdanov1994,PRBButenko2010}. 
One might have thought that the perpendicular field $H$ and easy-axis anisotropy would both favor
a skyrmion, all of whose spins are pointing up far from the center, but then the FM state is even more favorable.

\begin{figure}
\centering
\centerline{\includegraphics[width=7.5cm]{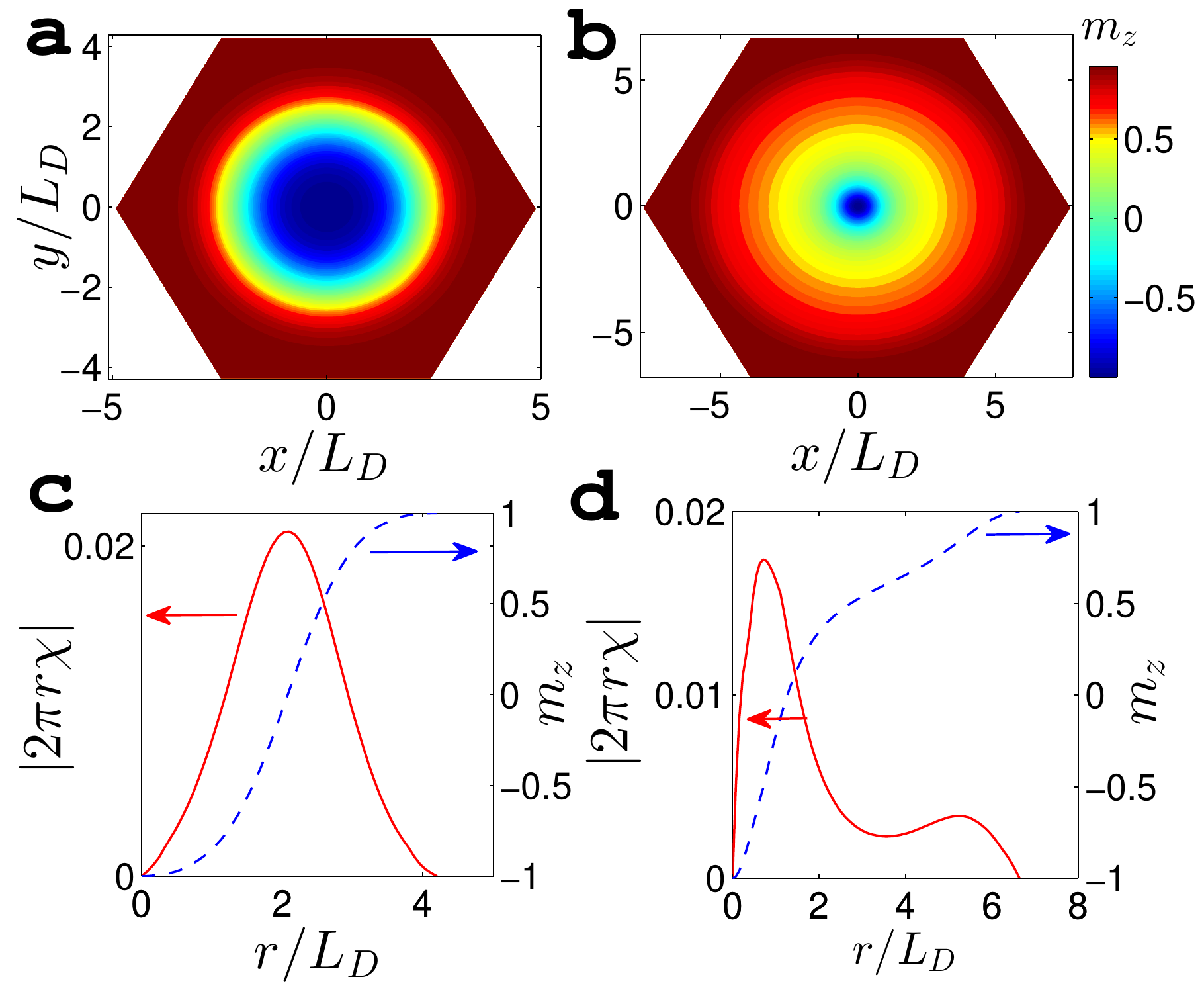}}
\caption{ {\bf Internal structure of skyrmion}: Skyrmion core structure from circular-cell calculation with $D/J=0.01$ and $A_cJ/D^2=1/2$. 
Here $L_D = (J/D)a$ where $a$ is the microscopic lattice spacing.
(a, b): False color plots of $m^z$ (shown in color bar).
(c, d): Angle-averaged topological charge density $|2\pi r \chi(r)|$ and $m^z(r)$ (right axes). 
Left panels (a) and (c) correspond to easy-axis anisotropy $AJ/D^2=-0.5$ and $HJ/D^2=0.28$. 
The skyrmion core is conventional with a single peak in the topological charge density.
Right panels (b), (d) are for easy-plane anisotropy $AJ/D^2=1.35$ and $HJ/D^2=1.96$.  
Here the core has a large `transition' region (yellow-orange) from down (center) to up (boundary) in $\mb{m}$
leading to an unusual two-peak structure for $|2\pi r\chi|$.}
\label{fig.skyrmion-structure}
\end{figure}

The remarkable result in Fig.~\ref{fig.PhaseDiagram}(b) is that the SkX phase is much more robust for 
easy-plane anisotropy ($A \!>\!0$). We can understand this as follows.
The twisted spins in the skyrmion lower the DM contribution to the free energy as compared to a ferromagnetic configuration.
Furthermore, the skyrmion is a better compromise between easy-plane anisotropy and a field along $\hat{z}$ than is a spiral configuration.
Thus, the large SkX region in the phase diagram is more or less oriented around $H=2A$,
the dashed line in Fig.~\ref{fig.PhaseDiagram}(b) that separates the `tilted FM' from easy-axis FM.
 
The internal structure of a skyrmion gives further insight into the stability of the SkX phase.
In Fig.~\ref{fig.skyrmion-structure} we plot $m^z(\mb{r})$ and the (angular averaged) topological charge density 
$|2\pi r\chi(r)|$, where $\chi(\mb{r})=[\mb{m}\!\cdot\!\partial_x\mb{m}\!\times\!\partial_y\mb{m}]/4\pi$.
For the easy-axis case the skyrmion core shows a conventional structure with a single peak in $|2\pi r\chi|$
 in Fig.~\ref{fig.skyrmion-structure}(a,c). In contrast, easy-plane anisotropy can lead to 
 a non-trivial core with a double peak in $|2\pi r\chi(r)|$; see Fig.~\ref{fig.skyrmion-structure}(b,d). As the spins twist from down at the
center ($\theta(0)\!=\!\pi$) to all up ($\theta(R)\!=\!0$)  at the boundary, it is energetically favorable to have an 
extended region where $\theta(r) \simeq  \theta_\mr{tilt}$
(see Appendix  ~\ref{App:CellSize}), the best compromise between the field and easy-plane anisotropy.
As a result, $|2\pi r\chi|$ shows a two-peak structure in the topological charge density.

\subsection{Phase transitions} We next describe the various phase transitions within our variational framework. 
The transitions between the spiral state and FM or SkX states are first order, with a crossing of energy levels,
as is the SkX to tilted FM transition for $H < 2A$.
These are all denoted by double lines in Fig.~\ref{fig.PhaseDiagram}.
The SkX to easy-axis FM transition for $H > 2A$ (denoted by a single line) is also first order in our numerics, but 
with the unusual feature that the optimal SkX unit cell size diverges at this transition; see Appendix ~\ref{App:CellSize}. 
Another interesting feature of Fig.~\ref{fig.PhaseDiagram} are the 
reentrant transitions from FM $\rightarrow$ SkX $\rightarrow$ FM for $AJ/D^2\apgt 1$.

%
\section{Microscopic analysis of exchange, DM and anisotropy}
We next present a microscopic, quantum mechanical derivation of the phenomenological free energy \eqref{eq.GL}, and show that 
the parameter regime of interest arises naturally for three very different exchange mechanisms -- superexchange, RKKY and double exchange -- 
in the presence of SOC.
Moriya's original paper~\cite{PRMoriya1960} considered antiferromagnetic superexchange with SOC,
further elaborated in a way relevant to our analysis in ref.~\onlinecite{PRLShekhtman1992}. 
The RKKY interaction with SOC was first discussed for spin-glasses~\cite{PRLFert1980} and the relation between 
DM and anisotropy was analyzed~\cite{PRBImamura2004} in the context of quantum dots. 
Double exchange ferromagnets with SOC were analyzed in our recent work~\cite{NaturePhysBanerjee2013}. 
In all these cases, it was found by explicit calculations that $A_c|J|/D^2=1/2$ (in the notation of this paper).

We sketch here a ``unified" way of thinking about these very different problems. 
We begin by summarizing the idea of our approach, before presenting details of
the derivation. We start with a microscopic Hamiltonian 
 $\mc{H}=\mc{H}_0+\mc{H}_{int}$ for electrons, where $\mc{H}_0$ is quadratic piece
 with the electronic kinetic energy and the SOC term, while $\mc{H}_{int}$ describes the
interactions that give rise to magnetism.
We then proceed as follows. First, in order to derive an effective low-energy Hamiltonian 
that describes magnetism, we consider a two-site problem, which is adequate for
situations where the magnetism can be ultimately described by pair-wise interactions between spins.
Next, we transform the original electrons by an $SU(2)$ rotation which ``gauges away'' the spin-orbit coupling. 
We can easily solve our problem in the rotated basis, since it now looks like the standard quantum magnetism problem 
without SOC.
Finally we transform back to the original physical electron basis and thus obtain, in addition to the exchange 
interaction, the DM and anisotropy terms. Since all three originate from a single interaction in the rotated basis,
there is a simple relation between the coefficients of the exchange, DM and anisotropy terms,
namely $A_c|J|/D^2=1/2$ (in the regime of weak SOC).

The quadratic piece in the microscopic Hamiltonian $\mc{H}=\mc{H}_0+\mc{H}_{int}$ 
is of the form
\begin{eqnarray}
\mc{H}_0=\!-t\!\sum_{<ij>,\alpha}\!c_{i\alpha}^\dag c_{j\alpha}
\!-\!i\lambda\sum_{<ij>,\alpha\beta}\!\boldsymbol{\sigma}_{\alpha\beta}.\hat{\mb{d}}_{ij}c_{i\alpha}^\dag c_{j\beta}+\mr{h.c.} \label{eq.Hkin}
\end{eqnarray}
Here the operator $c^\dagger_{i\alpha}$ ($c_{i\alpha}$) creates (destroys) an electron with spin $\alpha$ at a site $\mb{r}_i$
on a 2D square lattice. We consider for simplicity hopping $t$ between nearest-neighbors $<ij>$,
though our analysis can be easily generalized to a more general dispersion.
Here $\boldsymbol{\sigma}$ are Pauli matrices and $\lambda$ is the strength of the Rashba SOC
with $\hat{\mb{d}}_{ij}=\hat{\mb{z}}\times \mb{r}_{ij}/|\mb{r}_{ij}|$  with $\mb{r}_{ij}=\mb{r}_i-\mb{r}_j$. 
 
The interaction $\mc{H}_{int}$ can be chosen to model several different situations. 
(i) Hubbard repulsion $\mc{H}_{int}=U\sum_in_{i\uparrow}n_{i\downarrow}$ with $U\gg t$ at half-filling, 
gives rise to antiferromagnetic (AF) superexchange with SOC. 
Here $n_{i\alpha}=c^\dagger_{i\alpha}c_{i\alpha}$ is the electron number operator. 
(ii) Coupling of conduction electrons with a lattice of local moments 
$\mb{S}_i$ via $\mc{H}_{int}=-J_\mr{H}\sum_i \mb{s}_i\!\cdot\!\mb{S}_i$ leads to Zener double-exchange with SOC.
Here $\mb{s}_i\!=\!(1/2)\sum_{\alpha\beta}c_{i\alpha}^\dagger\boldsymbol{\sigma}_{\alpha\beta}c_{i\beta}$ and the Hund's coupling
$J_H\gg t$. 
(iii) The $\mc{H}_{int}$ of (ii) with a Kondo coupling $|J_\mr{K}|\ll t$ leads to an 
RKKY interaction between moments mediated by electrons with SOC. 
As explained above, in all three cases, the effective Hamiltonian can be derived by considering pairwise interaction between spins. 
We discuss (i) and (ii) below and relegate the RKKY case (iii) to Appendix \ref{App.RKKY}.
 
To derive an effective low-energy Hamiltonian, we consider a \emph{two-site} problem with nearest 
neighbor sites $1$ and $2$ and rewrite $\mc{H}_0$ for these sites as 
\begin{equation}
\mc{H}_0=-\tilde{t}\sum_{\alpha\beta}(c^\dag_{1\alpha}[e^{i\vartheta\boldsymbol{\sigma}.\hat{d}_{12}}]_{\alpha\beta}c_{2\beta}+\mr{h.c.})
\end{equation}
with $\tilde{t}=\sqrt{t^2+\lambda^2}$ and $\tan\vartheta=\lambda/t$.
Next we gauge away the SOC with $SU(2)$ rotations on the fermionic operators at the two sites, via 
$\tilde{c}_{1\alpha}=[e^{-i(\vartheta/2)\boldsymbol{\sigma}.\hat{d}_{12}}]_{\alpha\beta}c_{1\beta}$ and 
$\tilde{c}_{2\beta}=[e^{i(\vartheta/2)\boldsymbol{\sigma}.\hat{d}_{12}}]_{\alpha\beta}c_{2\beta}$. 
Under this transformation the non-interacting part simply becomes $\mc{H}_0=-\tilde{t}\sum_{\alpha}\tilde{c}^\dag_{1\alpha}\tilde{c}_{2\alpha}$,
as if there were no SOC (which is, of course, hidden in the parameter $\tilde{t}$ and the transformed fermion operators).

We next discuss how the interaction terms transform under this rotation. For the superexchange case (i) 
we find $\mc{H}_{int}=U\sum_{i=1,2}\tilde{n}_{i\uparrow}\tilde{n}_{i\downarrow}$ 
where $\tilde{n}_{i\alpha}$ is the number operator for rotated fermions. 
For the double exchange case (ii) we find $\mc{H}_{int}=-J_\mr{H}\sum_{i=1,2}\tilde{\mb{s}}_i.\tilde{\mb{S}}_i$ 
with $\tilde{\mb{s}}_i$ the spin operator for the rotated fermions and the local moments are
transformed as follows.
$\tilde{\mb{S}}_{1}$ is given by
\begin{equation}
\tilde{\mb{S}}_1\!=\!\cos{\vartheta}\mb{S}_1-\sin{\vartheta}(\mb{S}_1\times \hat{d}_{12})+(1-\cos{\vartheta})(\mb{S}_1.\hat{d}_{12})\hat{d}_{12}
\end{equation} 
while $\tilde{\mb{S}}_2$ can be obtained by replacing $1\rightarrow 2$ and $\vartheta\rightarrow -\vartheta$ in the above equation.
A similar relation holds between $\tilde{\mb{s}}_{1}(\tilde{\mb{s}}_{2})$ and $\mb{s}_{1}(\mb{s}_{2})$. 

The transformed Hamiltonian $\mc{H}$ for the two-site problem in terms of rotated fermions is exactly the same as the model without SOC. 
As a result, in case (i) superexchange between $\tilde{\mb{s}}_1$ and $\tilde{\mb{s}}_2$ has the usual form, $J_\mr{AF}\tilde{\mb{s}}_1.\tilde{\mb{s}}_2$ 
with $J_\mr{AF}=4\tilde{t}^2/U$. We now transform back to the original spin variables and write the superexchange Hamiltonian
as a sum over all near-neighbor pairs to obtain
$\mc{H}_{\rm SE}\!=\!J_{\rm AF}\sum_{<ij>}\mb{s}_i\!\cdot\!\mc{R}(2\vartheta\hat{{\bf d}}_{ij})\mb{s}_j$. 
Here $\mc{R}(2\vartheta\hat{{\bf d}}_{ij})$ is the orthogonal matrix corresponding to a rotation by angle $2\vartheta$ about $\hat{\mb{d}}_{ij}$.

The same argument applies for (ii) double-exchange case. Following Anderson-Hasegawa\cite{PRAnderson1955}, in the limit $J_\mr{H}\rightarrow \infty$ and for large (classical) spins $\tilde{\mb{S}}$ of the local moments, 
the effective exchange is $-J_\mr{F}\sqrt{1+\tilde{\mb{S}}_1.\tilde{\mb{S}}_2/2S^2}$. 
Here $S$ is the magnitude of the spin and $J_{\rm F} = \kappa\tilde{t}$ with $\kappa$ a constant that depends on the density of itinerant electrons. 
Again going back to $\mb{S}$'s and summing over pairs, one obtains the double-exchange Hamiltonian 
$\mc{H}_{\rm DE}=-J_{\rm F}\sum_{<ij>}\left[(1+ \mb{S}_i\!\cdot\!\mc{R}(2\vartheta\hat{{\bf d}}_{ij})\mb{S}_j/{S^2})/2\right]^{1/2}$.

At low-temperatures, the effective spin model for both cases (i) and (ii) can be written in a common form (after expanding
the square-root in case (ii) and a sublattice rotation in case(i)). 
We get 
\begin{eqnarray}
\mc{H}_{eff}&=&-J\sum_{i,\mu}\mb{S}_i\!\cdot\!\mb{S}_{i+\hat{\mu}}-A_c\sum_i (S^y_iS^y_{i+\hat{x}}+S^x_iS^x_{i+\hat{y}})\nonumber \\
&&-D\sum_i[\hat{y}\!\cdot\!(\mb{S}_i\times \mb{S}_{i+\hat{x}})-\hat{x}\!\cdot\!(\mb{S}_i\times \mb{S}_{i+\hat{y}})],\label{eq.H_eff}
\end{eqnarray} 
where $\hat{\mu}=\hat{x},\hat{y}$. Here $J=\tilde{J}\cos{2\vartheta}$ with $\tilde{J}=J_\mr{AF}$ for super-exchange 
and $\tilde{J}=J_\mr{F}$ for double-exchange. The SOC-induced terms are the DM term with 
$D=\tilde{J}\sin{2\vartheta}$ and the compass anisotropy $A_c=\tilde{J}(1-\cos{2\vartheta})$.
Since $\tan{\vartheta}=\lambda/t \ll 1$, we get the microscopic result $A_cJ/D^2 \simeq 1/2$.

It is straightforward to derive the continuum free energy \eqref{eq.GL} from the lattice model \eqref{eq.H_eff}.
The only term in \eqref{eq.GL} that does not come from \eqref{eq.H_eff} is the phenomenological anisotropy
$A_s(m^z)^2$ arising from single-ion or dipolar shape anisotropy~\cite{RepProgPhysJohnson1996}.
In some cases, a simple estimate of dipolar anisotropy is much smaller than the compass term~\cite{NaturePhysBanerjee2013}. 
For moments with $S < 2$, the single-ion anisotropy vanishes~\cite{BookYoshida}. 
For larger-$S$ systems, the single-ion anisotropy is non-zero and can even be varied using strain~\cite{PRBButenko2010}.
However, in no case can we ignore compass anisotropy, since its contribution to the energy is comparable to DM, as
already emphasized.

%
\section{Discussion}
We now discuss two important questions: (a) the applicability of our results with Rashba SOC to quasi-2D systems or films with finite thickness,
and (b) the differences between the broken surface or $z$-inversion, which has been our primary focus here, and broken bulk inversion.

First, let us consider quasi-2D systems made of materials that do {\it not} break bulk-inversion. Chiral interactions then arise only from
Rashba SOC.  It might seem, at first sight, that the effects of surface-inversion breaking would be restricted to very thin, possibly monolayer, 
samples. However, it is known in the semiconductor literature that Rashba SOC can be very strong even in films of thickness
of order a micron~\cite{KatoNature2003} due to strain effects. Thus we believe that the 2D results described in this paper
are not restricted as such to monolayer materials. The spatial variation of the local magnetization ${\bf m}({\bf r})$ will be
translationally invariant in the z-direction, and the SkX phase will continue to show the large region of
stability for in-plane anisotropy shown in Fig.~\ref{fig.PhaseDiagram}.

In systems with broken bulk inversion, a cone phase~\cite{PRBWilson2014} overwhelms both the SkX and FM
in the easy-plane anisotropy regime. The cone phase, with spin texture varying along the field axis, gains energy
due to a DM term with ${\bf D}_{i,i+\hat{\bf z}} \parallel \hat{\bf z}$. Such a term does not exist in 2D or even in quasi-2D systems
with Rashba SOC, where the DM vector $\mb{D}_{ij} \sim \widehat{\mb{z}}\times\widehat{\mb{r}}_{ij}$ lies in the $xy$-plane.

Our phase diagram is thus completely different from that of ref.~\onlinecite{PRBWilson2014} for the case of in-plane anisotropy.
Ref.~\onlinecite{PRBWilson2014} considers bulk-inversion breaking with a $\mb{m}\cdot(\nabla\!\times\!\mb{m})$ DM term and finds
a stable cone-phase for $A>0$. We, on the other hand, consider surface-inversion breaking with Rashba SOC leading to the DM term of
eq.~(\ref{Rashba-DM}) and find a large region where the SkX is stable for $A>0$.

An interesting question arises for a quasi-2D system, such as a thin film, made of a material that breaks
bulk inversion. Now one has to take into account both Dresselhaus and Rashba terms arising from bulk and surface
inversion breaking, respectively. We will show elsewhere~\cite{RowlandUnpublished} that by tuning the relative strengths
of Rashba to Dresselhaus SOC one can continuously interpolate between the results presented here (only surface inversion broken) and those
of ref.~\onlinecite{PRBWilson2014} (only bulk inversion broken) with interesting evolution of skyrmion chirality from hedgehog-like
to vortex-like. Interestingly, the data in Fig.~\ref{fig.PhaseDiagram} of ref.~\onlinecite{PRLLi2013} show a SkX phase 
in epitaxial MnSi thin films for thickness $\lesssim (J/D)a$.

\section{Conclusions}
We have shown enhanced stability of skyrmions in 2D for Rashba SOC when the effective anisotropy is easy-plane.
The compass term $A_c$ is intrinsically easy-plane and we suggest that experiments should look for 2D systems with
suitable single-ion anisotropies $A_s$, or ways to tune it, e.g., using strain, so as to enhance the SkX region.
In the future, it would be interesting to study the finite temperature phase diagram for 2D systems with easy-plane anisotropy,
and to understand electronic properties, like the topological Hall effect and non-Fermi liquid behavior in this regime.

\begin{acknowledgments}
We gratefully acknowledge discussions with
C. Batista, R. Kawakami, D. Khomskii, C. Pfleiderer, and especially S-Z. Lin,
and the support of DOE-BES DE-SC0005035 (S.B., M.R.), 
NSF MRSEC DMR-0820414 (J.R.) and NSF-DMR-1006532 (O.E.). 
M.R. acknowledges the hospitality of the Aspen Center for Physics.

\end{acknowledgments}

\appendix

\section{Variational calculation}\label{App:VarCaln} 

We consider the FM, spiral and SkX phases in turn.
We use $A = A_c+A_s$ as the effective anisotropy, and
omit additive constants in the energy, which are common to all phases.

\bigskip
{\bf FM:} The energy for the FM state evaluated from eq.~\eqref{eq.GL} is ${F}_\mr{FM}=A(m^z)^2-Hm^z$ with minimum ${F}_\mr{FM}=-H^2/4A$ for $H\leq 2A$ and ${F}_\mr{FM}=A-H$ 
for $H > 2A$ 
The corresponding magnetizations are $m^z=H/2A$ and $m^z=1$ respectively.

For the $A\!>\!0$ FM state, the easy-plane anisotropy competes with the field along $\hat{z}$ so
that the magnetization points at an angle $\theta_\mr{tilt}=\cos^{-1}(H/2A)$ with respect to the $z$-axis for $H\leq 2A$ and 
eventually aligns with the field for $H>2A$. We denote the FM state for $H\leq 2A$ as the `tilted FM'. The dashed line $H = 2A$
in Fig.~\ref{fig.PhaseDiagram}(b) separates the field-aligned FM from the tilted FM.

\bigskip
{\bf Spiral:} 
The simplest zero-field variational ansatz~\cite{NaturePhysBanerjee2013}
yields a FM ground state for $|A|J/D^2>1$. When $|A|J/D^2<1$, the $H\!=\!0$ ground state is a coplanar spiral with spins lying in a plane perpendicular to the $xy$-plane: $\mb{m}(\mb{r})=\sin(\mb{Q}_0.\mb{r})\hat{\mb{Q}}_0+\cos(\mb{Q}_0.\mb{r})\hat{z}$ 
with $\mb{Q}_0=(D/J)(\cos\varphi\hat{x}+\sin\varphi\hat{y})$.

We extend the simple spiral above to incorporate more general 1D modulation described by 
$
\mb{m}_{\rm spiral}(\mb{r})=\sin[\theta(\mb{Q}_0.\mb{r})]\hat{\mb{Q}}_0+\cos[\theta(\mb{Q}_0.\mb{r})]\hat{\mb{z}},
$ 
where $\theta$ varies only along $\hat{\mb{Q}}_0$, chosen to be $\hat{x}$ without loss of generality.
In contrast to the linear variation in the simplest ansatz, here $\theta(x) $ is an arbitrary function 
with $\mb{m}(x+R)=\mb{m}(x)$ where $R$ is the period. We numerically minimize \eqref{eq.GL} 
with the variational parameters $\theta(x)$ and $R$. 
This more general 1D periodic modulation stabilizes the spiral relative to FM beyond $|A|J/D^2=1$ to 
$\simeq 1.25$ at $H=0$; see Fig.~\ref{fig.PhaseDiagram}(b). 

With the general 1D periodic modulation, 
$\mb{m}(x)=\sin{\theta(x)}\hat{x}+\cos{\theta(x)}\hat{z}$, the free energy is
\begin{eqnarray}
{F}_\mr{sp}&=&\frac{1}{R}\int_0^R dx \left[(J/2)\left(\partial_x \theta\right)^2-D\partial_x \theta\right.\nonumber \\
&&\left.+A\cos^2{\theta}-H\cos\theta\right],
\end{eqnarray}
where $\partial_x\theta=(\partial \theta/\partial x)$. We use conjugate gradient minimization with respect to the size $R$ and the function
$\theta(x)$ which is discretized on a 1D grid. 
We use the periodic boundary condition $\theta(R)=\theta(0)+2\pi n$ where $n$ is an integer. 
This form allows for a spiral solution with a net magnetization $m^z$ in the presence of a perpendicular magnetic field.

The more restrictive (linear) variational ansatz $\theta(x)=2\pi (x/R)$ is equivalent to the previously studied case~\cite{NaturePhysBanerjee2013} and is analytically tractable.
In this case the energy of the spiral can be easily evaluated by minimizing with respect to $R$. 
This gives the spiral pitch $R=R_\mr{sp}=2\pi(J/D)$ and the energy ${F}_\mr{sp}=-D^2/2J + A/2$. 

\bigskip
{\bf Skyrmion crystal:} We have discussed in the main paper the method used to construct a hexagonal SkX solution using the 
circular cell ansatz with rotationally symmetric form of eq.~\eqref{eq.Skyrmion} in the text. 

To qualitatively understand the stability of SkX over FM and spiral states, 
one can use a simple linear ansatz $\theta(r)=\pi(1-r/R)$ and minimize the energy by choosing an optimal $R$.
This leads to the solution $R_\mr{sk}\approx \pi J/D$ for the optimal skyrmion cell size, with the energy given by
\begin{eqnarray}
F_\mr{sk}&=&\frac{-\pi^2}{2[\pi^2+\gamma+\log(2\pi)\!-\!\mr{Ci}(2\pi)]}{ D^2 \over J} + \frac{A}{2} - \frac{4}{\pi^2}H \nonumber\\
&\simeq& - 0.4009{ D^2 \over J} + \frac{A}{2} - \frac{4}{\pi^2}H
\end{eqnarray}   
Here $\mr{Ci}(x)=-\int_x^\infty dt\ {\cos{t}}/{t}$ is the cosine integral and $\gamma$ is the Euler constant. 
The result for $F_\mr{sk}$ makes it clear that SkX gains energy from both DM and Zeeman terms.

For the more general $\theta(r)$ variation within the circular cell ansatz, we need to numerically minimize
\begin{eqnarray}
{F}_\mr{sk}&=&\frac{2}{R^2}\int_0^R r dr\left[e_J+e_D+e_C+e_S-H\cos\theta\right].
\end{eqnarray}
with
\begin{eqnarray*}
e_J&=&\frac{J}{2}\left[\left(\frac{\partial \theta}{\partial r}\right)^2 + {\sin^2\theta \over r^2}\right]\nonumber\\
e_D&=&-D\left[\frac{\partial \theta}{\partial r}+\frac{\sin2\theta}{2r}\right] \nonumber \\
e_C&=&A_c\cos^2\theta+\frac{A_c}{8}\left[\cos\theta\left(\frac{\partial \theta}{\partial r}\right)-\frac{\sin\theta}{r}\right]^2\nonumber\\
e_S&=&A_s\cos^2\theta
\end{eqnarray*}
We need to find the optimal cell size $R$ and optimal values of $\theta(r)$, which we discretize on a 1D grid in the radial direction. 
We have carried out 1D conjugate gradient minimization using Mathematica on a laptop,
using grids of up to 250 points.

\begin{figure}[htps]
\begin{center}
\includegraphics[width=7.5cm]{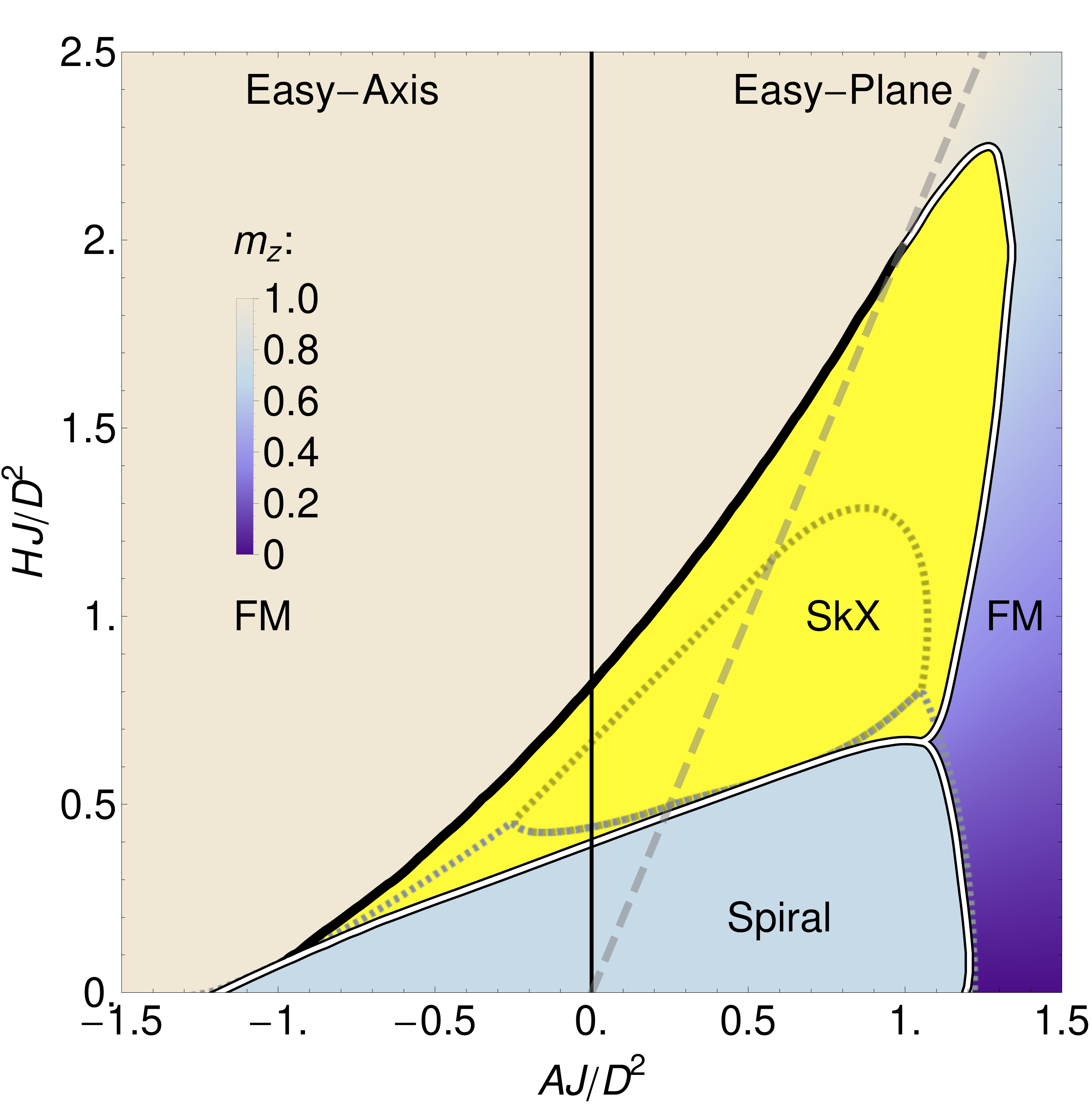}
\end{center}
\caption{{\bf $T\!=\!0$ anisotropy-field phase diagram from linear ansatz and 2D square cell calculation}: 
The phase diagram shown here is obtained as a result of a full 2D 
variational calculation, as distinct from the effectively 1D variational calculation shown in Fig.~\ref{fig.PhaseDiagram}(b).
The symbols and parameters used are exactly the same as described in the caption for Fig.~\ref{fig.PhaseDiagram}(b).
Note that the 2D square cell calculation and the 1D variational calculation, although quite different in their computational complexity, 
nevertheless lead to essentially identical results for the overall phase diagram.
The dotted boundaries shown here are obtained from the simplest `linear' variational ansatz for SkX described in text.
}
\label{fig.PhaseDiagram_SquareCell}
\end{figure}

\begin{figure}[htps] 
\centering
\centerline{\includegraphics[height=7.5cm]{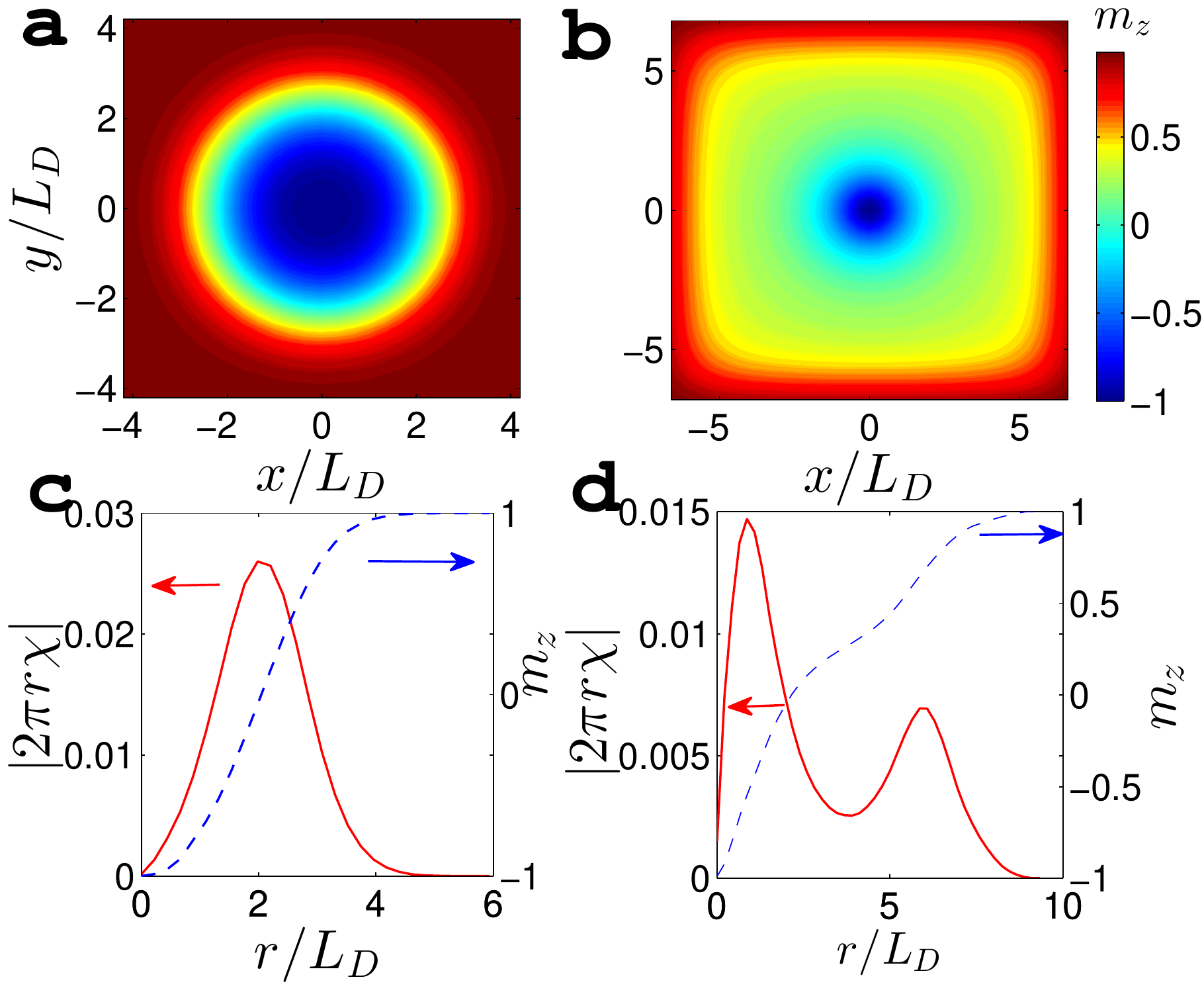}}
\caption{ {\bf Internal structure of skyrmion from 2D square cell calculation}: Skyrmion core structure 
with $D/J=0.01$ and $A_cJ/D^2=1/2$ obtained from a full 2D variational calculation, which
should be compared with circular-cell results shown in Fig.~\ref{fig.skyrmion-structure}.
(a, b): False color plots of $m^z$ (shown in color bar).
(c, d): Angle-averaged topological charge density $|2\pi r \chi(r)|$ and $m^z(r)$ (right axes). 
Left panels (a) and (c) correspond to easy-axis anisotropy $AJ/D^2=-0.5$ and $HJ/D^2=0.3$. 
Right panels (b), (d) are for easy-plane anisotropy $AJ/D^2=1.2$ and $HJ/D^2=1.1$.  
Note that the parameters used here are slightly different from those used in Fig.~\ref{fig.skyrmion-structure},
however the nontrivial structure of the skyrmion core in the easy-axis case is qualitatively similar to that
in the circular cell calculations.}
\label{fig.skyrmion-structure_SquareCell}
\end{figure}

\begin{figure}[htps]
\begin{center}
\includegraphics[width=7.5cm]{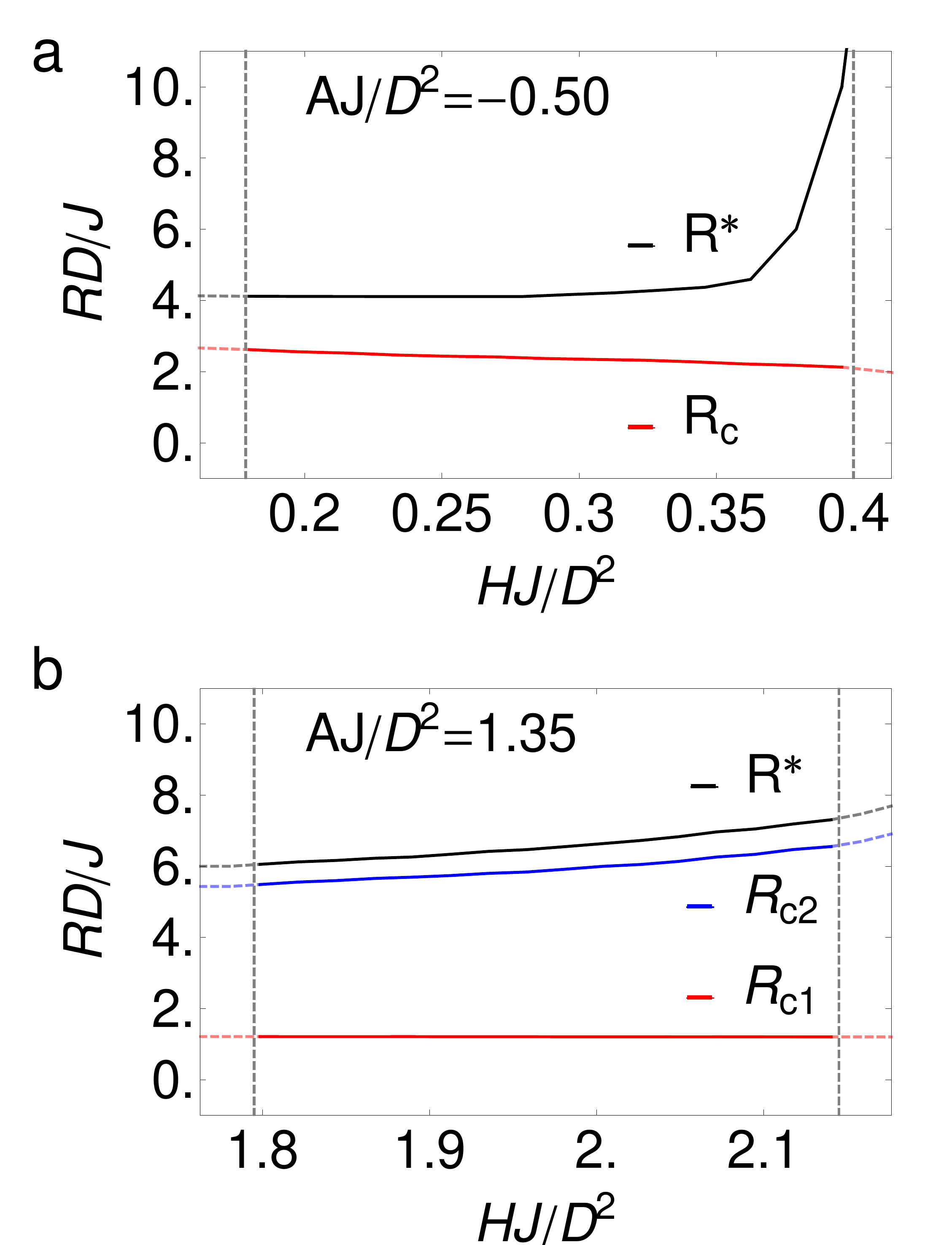}
\end{center}
\caption{{\bf Skyrmion length scales}: 
Plots of the $H$-dependence of the optimal skyrmion 
cell radius $R^*$ and the core radii defined by the location of the maxima of $|dm^z/dr|$. 
For the ansatz of eq.~\eqref{eq.Skyrmion} in the text, $dm^z/dr=|2\pi r\chi(r)|$.
(a) In the easy-axis region, both $R^*$ and the core radius $R_c$
are finite at the spiral-to-SkX phase boundary, but $R^*$ diverges
while $R_c$ remains finite at the SkX-to-FM transition.
The vertical dashed lines indicate phase transitions from the SkX to the
spiral state (at small $H$) and to the FM (at large $H$).
(b) In the easy-plane region, there are two core radii corresponding to the two maxima 
in $|2\pi r\chi(r)|$. These inner and outer core radii $R_{c1}$ and $R_{c2}$,
and the cell radius $R^*$, all remain finite at the two phase transitions
out of the SkX phase.
Here the vertical dashed lines indicate SkX-FM transitions.
}
\label{fig.Radius}
\end{figure}

\section{2D Minimization}\label{App:2DMin} 

To check the validity of the circular cell ansatz,
we have also performed a full 2D minimization by discretizing the GL functional (1) over a square grid. 
For the 2D calculation, we used up to $100 \times 100$ grids with polar and azimuthal angles $(\theta(\mb{r}),\phi(\mb{r}))$ 
of $\mb{m}(\mb{r})$ at each grid point as variational parameters. The 2D conjugate gradient
calculations are done using a Numerical Recipes~\cite{SPressNR} subroutine in C on a local cluster of computers.
This 2D minimization is much more computationally intensive than the 1D calculation for the
circular cell ansatz.

The 2D square cell result shown in Fig.~\ref{fig.PhaseDiagram_SquareCell} for the phase diagram is essentially
the same as that obtained from the circular cell calculation; see Fig.~\ref{fig.PhaseDiagram}(b). 
We show in Fig.~\ref{fig.skyrmion-structure_SquareCell} the internal structure of the skyrmion as calculated 
from the full 2D square-cell minimization. This figure should be compared with the 
results from a circular cell calculation in Fig.~\ref{fig.skyrmion-structure}.
Note that the parameters used here are slightly different from those used in Fig.~\ref{fig.skyrmion-structure},
however the nontrivial structure of the skyrmion core in the easy-axis case -- the two-peak structure in 
the topological charge density $|2\pi r \chi(r)|$ --
is qualitatively similar to that in the circular cell calculations.

\section{Skyrmion cell size and core radius}\label{App:CellSize} 

It is conventional to define the `core radius' of a skyrmion from the maximum of $|dm^z/dr|$. 
For the rotationally symmetric ansatz, eq.~\eqref{eq.Skyrmion} in the main text, $dm^z/dr=|2\pi r\chi(r)|$.

In Fig.~\ref{fig.Radius} 
we show the optimal skyrmion cell size $R^*$ and core radii as a function of field for
(a) easy-axis anisotropy with $AJ/D^2=-0.5$ and (b) easy-plane anisotropy with $AJ/D^2=1.35$. 
As described in the main paper, and shown in Fig.~\ref{fig.skyrmion-structure}, 
there is only one length scale associated with skyrmion core size for the easy axis case, where as two-length scales  
appear for the easy-plane side, near the re-entrant region of the SkX phase diagram. 

The inner core radius, denoted $R_c$ and $R_{c1}$ in Fig.\ref{fig.Radius} (a) and (b) respectively, 
is found to be essentially constant as a function of $H$; it is
fixed by the competition between ferromagnetic and DM terms to a value of order $J/D$.
On the other hand, the optimal skyrmion cell size $R^*$ can have non-trivial variation with $H$.
For instance, in Fig.\ref{fig.Radius}(a) we see that $R^*$ diverges at the 
phase boundary between the SkX and out-of-plane FM in the easy-axis case. 
In this case the skyrmion spins change from down to up on the length scale $R_c$, and then remain up out to $R^*$. 
We next discuss the implications
of the divergence of $R^*$ for the nature of the phase transition.

\section{Phase transition from SkX to easy-axis FM}\label{App:PhaseTrans}

The optimal cell size diverges with
$R^* \rightarrow \infty$ as $H \rightarrow H_c^-$ as one approaches the phase boundary (shown as a full line
in the phase diagrams in Fig.~\ref{fig.PhaseDiagram} and Fig.~\ref{fig.PhaseDiagram_SquareCell}). We note that $R^*$,
which determines the wave vector for the SkX Bragg spots, is not a correlation length. 
Nevertheless, a divergent length scale raises the question:
Is this transition continuous or first order? 

To determine the order of the transition 
we need to know whether the SkX and FM energy densities approach each 
other with zero relative slope (continuous transition) or a finite slope difference (first-order transition),
as a function of field $H$ (at fixed anisotropy).
While this is hard to nail down numerically without a careful finite size scaling analysis, 
the following argument is instructive.

We focus on the energy density difference between the SkX and FM
\begin{equation}
\Delta {\cal E} \equiv {\Delta E / L^2} =  (E_{sk} - E_{fm})/ L^2
\end{equation}
in the thermodynamic limit $L \rightarrow\infty$. We can write
the SkX energy density as
\begin{equation}
{E_{sk}\over L^2} = {\cal E}_c \left(R_c\over R^*\right)^2 + {\cal E}_{fm}\left[1 - \left(R_c\over R^*\right)^2 \right].
\end{equation}
where ${\cal E}_c$ is the energy density of the skyrmion core of size $R_c$ and ${\cal E}_{fm}$ is that of the FM. 
We thus obtain
\begin{equation}
\Delta {\cal E} = \left[{\cal E}_c - {\cal E}_{fm}\right] \left(R_c / R^*\right)^2.
\label{eq:deltaE}
\end{equation}

Approaching the transition $h = (H-H_c)/H_c \rightarrow 0^-$,
the only singular quantity in eq.~(\ref{eq:deltaE}) is 
\begin{equation}
R^* \sim h^{-\nu^*}.
\end{equation}
All other quantities on the RHS of eq.~(\ref{eq:deltaE}) are smooth functions of $h$.
Thus we obtain
\begin{equation}
\Delta {\cal E} \approx F(h) h^{2\nu^*},
\end{equation}
where $F(h)$ is a smooth polynomial with $F(0) \neq 0$, since we do not expect
$({\cal E}_c - {\cal E}_{fm})$ to vanish in general for $h=0$.

The difference in the slopes of the energy density is then given by
\begin{equation}
{\partial \Delta {\cal E} \over \partial h} = F^\prime(h) h^{2\nu^*} + 2\nu^* F(h) h^{2\nu^* - 1}.
\label{eq:slope}
\end{equation}
For $h \rightarrow 0$, the first term vanishes, but the behavior of the second term
depends on $\nu^*$. 
If $\nu^* = 1/2$ we get a finite slope difference in eq.~(\ref{eq:slope}) and thus a first order transition from 
the energy perspective, despite a divergent length scale. 
However, if $\nu^* > 1/2$ then we would get a continuous transition.
Within a variational or mean field calculation one might expect $\nu^* = 1/2$, although
this is not entirely clear since $R^*$ is not a correlation length associated with an order parameter.

\section{RKKY interactions with SOC}\label{App.RKKY} 
Here we discuss the case of RKKY~\cite{SYoshidaRKKY} 
interaction between local moments embedded in a metallic host described by 
eq~(3) in the text. In this case, the magnetic exchanges~\cite{PRBImamura2004}, namely the isotropic, DM and compass, 
between two moments at $\mb{r}_1$ and $\mb{r}_2$ turn out to be 
$J_{12}=\tilde{J}(r_{12})\cos{2\vartheta_{12}}$, 
$D_{12}=\tilde{J}(r_{12})\sin{2\vartheta_{12}}$ and 
$A_{12}=\tilde{J}(r_{12})(1-\cos{2\vartheta_{12}})$. 
Here $\mb{r}_{12}=\mb{r}_2-\mb{r}_1$, $\vartheta_{12}=k_Rr_{12}$ 
with $k_R\equiv (\lambda/ta)$ and $\tilde{J}(r)\simeq -(J_K^2a^2/4\pi^2t)\sin{(2k_\mr{F}r)}/r^2$. 
This result is obtained \cite{PRBImamura2004} for $k_\mr{F}r_{12}\gg 1$, where $k_\mr{F}$ 
is the Fermi wavevector, and by approximating 2D tight-binding energy 
dispersion by a parabolic band, as appropriate for low-density of conduction electrons. 
Evidently, for $\lambda\ll t$ and $k_\mr{F}^{-1}\ll r_{12}\ll k_R^{-1}$, 
the ratio $A_{12}J_{12}/D_{12}^2\simeq 1/2$ is maintained. 
We consider a set of moments that are regularly distributed on a square lattice with 
a spacing $a$ such that the ratio $AJ/D^2\simeq 1/2$ for nearest-neighbor exchanges.
If we neglect longer-range part of the RKKY, then we obtain the effective spin Hamiltonian ~\eqref{eq.H_eff}
of the main text.

\bibliography{SkX-refs}

\end{document}